\documentclass[12pt, english]{article}
\usepackage[latin1]{inputenc}
\usepackage{graphics}
\usepackage{amsmath}
\usepackage{amssymb}
\usepackage{latexsym}
\usepackage{amsfonts}

\begin{document}

\hspace{4cm}

\begin{center}
{\large{\bf A precise formulation of the third law of thermodynamics}}
\end{center}

\vspace{2cm}

\begin{center}
  {\sc Walter F. Wreszinski} and {\sc Elcio Abdalla}\\
  Departamento de Física Matemática,\\[-0.2cm]
  Instituto de Física,
  Universidade de São Paulo, \\[-0.2cm]
  Caixa Postal 66318 - 05315-970 São Paulo - Brazil
\end{center}

\vspace{1cm}

\noindent{\small{
{\sc Abstract} ---
  The third law of thermodynamics is formulated precisely: all points
  of the state space of zero temperature $\,\Gamma_0\,$ are physically
  adiabatically inaccessible from the state space of a simple
  system. In addition to implying the unattainability of absolute zero
  in finite time (or ``by a finite number of operations''), it admits as
  corollary, under a continuity assumption, that all points of
  $\,\Gamma_0\,$ are adiabatically equivalent. We argue that the
  third law is universally valid for all macroscopic systems which
  obey the laws of quantum mechanics and/or quantum field theory. We also 
  briefly discuss why a precise formulation of the third law for black holes remains an open problem.

\vspace{2cm}



\newpage

\noindent{\bf 0 -  Introduction and Summary}
\vspace{0.3cm}

In a nice review on black holes and thermodynamics~[1], Wald
remarks that there are two completely independent statements referred
to as the third law. ``The first statement consists of the rather
vague claim that it is physically impossible to achieve $\,T=0\,$ for a
(sub) system.  The second statement, usually referred to as
`Nernst's  theorem', consists of the claim that $\,S(T)\rightarrow
0\,$ as $\,T\rightarrow 0$.'' Above, $\,T\,$ denotes the Kelvin
temperature and $\,S$, the entropy.

The trouble with the second statement is that it is violated by
several substances which display a residual entropy per particle$\, S_r\,$ as
$\,T\rightarrow 0$. The existence of a
nonzero $\,S_r\,$ has been explained in a quite general framework (of
quantum statistical mechanics) by the existence of a ground state
degeneracy which is exponentially large in the number of particles
(but the question of boundary conditions is very subtle)~[2], and the
actual value of $\,S_r\,$ for ice was calculated by Lieb in a seminal
paper, yielding a value in excellent agreement with experiment [3].
In ([1],[6]) it is stated that the second statement is also violated by black holes. We come back to this question briefly in section 3, but warn the reader that the problem remains open.

Finally, and perhaps most importantly, the second statement is
motivated by statistical mechanics, i.e., the formula
\[
S_r = k \ln W
\]
where $\,W\,$ denotes the ground state degeneracy: ``normally'' the
ground state is nonde-\linebreak generate,\footnote{It is amusing to
remark that, while Fermi considered an exponential degeneracy
implausible [4], Pauli simply states that $\,W=1\,$ is an assumption,
without further comment [5].}thus $\, W=1$, yielding $\, S_r =0\,$
(see [2] for a rigorous discussion of these issues in a class of
lattice models). However, statistical mechanics should not be invoked when one takes the standpoint that there exists a self-contained formalism - thermodynamics - of which the third law should be an integral part, and which has a certain ``universal applicability'' to a vast range of physical systems.

In this paper, we revisit the problem, and propose a precise
thermodynamic statement of the third law -- the unattainability of the
state space corresponding to zero temperature in a finite number of steps. This is done in section 1, on the basis of the  framework introduced by
Lieb and Yngvason [7] and of two assumptions, one of them of general
nature (assumption 1), the other, a continuity assumption (assumption
2), whose validity is expected for a wide class of systems (exceptions
are discussed ).  In section 1 we also show that the
third law implies a general result: all points of the state space corresponding to zero temperature are adiabatically equivalent (Theorem 1.1: Planck's formulation of Nernst's theorem). \textbf{All} substances, including those with $S_r\ne0$, should obey it. Examples of the latter are ice or $\,CH_3D\,$.
A startling recent example is artificial ``spin ice'', with $ S_r = 0.67 R\ln2$, close to the value $\, S_r = 0.71R\ln2\,$ determined for ice [8].

Section 2 briefly discusses the known proof [9] that, generally speaking, the
classical limit which holds for the free energy of a very wide class
of quantum spin systems by Lieb's classic paper [10], does not hold for
the entropy. This obstruction seems
to be at the heart of the fact that the third law does not hold for
classical systems in statistical mechanics. We also conjecture that, for classical spin systems and (non-relativistic) classical particle systems Assumption 2
(continuity) is universally \textbf{not} fulfilled, and illustrate it with two examples.

Section 3 is devoted to the conclusion, open problems and conjectures. Among the open problems, we briefly discuss the assertion in [6] that black holes do not stisfy the third law in the form of Planck's formulation (Theorem 1.1).

We refer to [11] for the derivation of the fundamental laws of
thermodynamics by a different route, viz., from nonequilibrium quantum
statistical mechanics, and to [12] for an alternative treatment of the
third law.

We believe that it is very significant that the sophisticated formulation of thermodynamics by Lieb and Yngvason in [7], suffices to provide a precise formulation of the third law, with Planck's formulation of Nernst's theorem as corollary. The required mathematics is elementary (see Theorem 1.1), but the clarification of the minimally required assumptions within a mathematically rigorous framework greatly increases our confidence that the third law is not a vague claim, but rather a universal law of nature, for systems which obey the laws of quantum mechanics and/or quantum field theory.

\vspace{1cm}

\noindent{\bf 1 -  Statement and Consequences of the Third Law}

\vspace{0.3cm}

We adopt the precise formulation of thermodynamics due to Lieb and
Yngvason~[3] and consider a simple system ([3], Sect. 3.2), with space
of states $\,\Gamma\,$ -- a subset of $\,R^n\,$ -- and points $\, X\in
\Gamma\;;\;\; X\prec Y\,$ means that $\,Y\,$ is adiabatically accessible
from $\,X\,$ in the sense of ([7], Def.  p.17) this definition does
agree with the usual motion based on processes taking place within an
``adiabatic enclosure'' ([7], Theorem 3.8). If $\, X\prec Y\,$ and 
$\, Y\prec X$, one says $\,X \stackrel{A}{\sim}\, Y\,$, i.e., $\,X\,$ and $\,
Y\,$ are {\em adiabatically equivalent\/}.  As remarked ([7], pg 17),
the word ``adiabatic'' is sometimes used to mean ``slow'' or
quasi-static, but not in the meaning ascribed to it by the authors:
``the explosion of a bomb in a closed container is an adiabatic
process''. Further examples include, of course, more commonly observed
processes, such as natural processes within an isolated compound
system after some barriers have been removed-mixing and chemical or
nuclear processes.

One writes [7]
\begin{equation}
X \prec\prec Y
\end{equation}
if $\, X\prec Y\,$ but $\,Y\nprec X $, i.e., $\,Y\,$ is
adiabatically accessible from $\,X\,$ ``in the real world''. If (1) is
true we shall say that $\,Y\,$ is physically adiabatically accessible
from  $\,X\,$ (for lack of a better name). Lieb and Yngvason remark that it is
possible to redo their axiomatization of thermodynamics using this
latter concept, and we shall adopt this strategy, quoting this
alternate version of their results. It is, however, very important to
stress that in [7] the states of a system are always {\it
equilibrium\/} states, although, as remarked there, the equilibrium
may depend on internal barriers.

In ([7], p.19) the second law was formulated as the {\it entropy
  principle\/}:

There is a real valued function on all states of all systems
(including compound systems), called entropy and denoted by $\,S\,$,
such that it is:

a) monotone:
\begin{equation}
X  \stackrel{A}{\sim}Y  \rightarrow S(X) = S(Y)
\end{equation}
\begin{equation}
X\prec\prec Y \rightarrow S(X) < S(Y)
\end{equation}

b) $S\,$ is additive and extensive. The latter property may be written
([7], 2.5)
\begin{equation}
S(tX) = tS(X)
\end{equation}
for each $\, t>0\,$, each state $\,X\,$, and each scaled copy $\,
tX\,$ (see [7], pg. 15). It is noteworthy that the entropy function
satisfying \ a.) and \ b.) was {\em constructed\/} in [7], and a
unicity property was shown there ([7], Theorem 2.2, for a single
system). Finally, at every point $\,X\,$ in the state space of a
simple system, $\,\Gamma\,$, a function, the temperature $\, T=T(X)$,
exists, such that thermal equilibrium $\, X_1 \stackrel{T}{\sim}
X_2\,$ (defined in [7], pg. 55) satisfies $\, X_1 \stackrel {T}{\sim}
X_2 \,$iff $\, T(X_1) = T (X_2)$, which is single-valued and unique
([7], theorem 5.1), given by
\begin{equation}
\frac{1}{T(X)} = \left(\frac{\partial S}
{\partial U}\right) (X)
\end{equation}
where the energy $\,U\,$ is defined in ([7], pg. 40).

By (5), $0< T(X) < \infty$, the points
\begin{equation}
T(X) =0
\end{equation}
and $\, T(X) = \infty\,$ being singular points of the state space
. Accordingly, we shall, for conciseness make the following
assumptions:

\vspace{0.3cm}

\noindent{\sc Assumption 1} \ The state space $\,\Gamma\,$ of the simple system is
parametrized by $\, \Delta \times I_T $, where $\, I_T \equiv
(0,T_0]$, $\Delta $ is a (finite dimensional) set,  ``$\times$'' denotes cartesian
product, and $\,T_0 >0$. We shall denote by $Z$  the points of the
set $\Delta$. 

\vspace{1cm}

\noindent{\sc Assumption 2} \ The entropy function $\, S = S(Z,T)\,$
may be uniquely extended to the closure $\,{\overline{I}_T} = [0,T_0]\,$
of $\, I_T\,$ by continuity:
\begin{equation}
S(Z,0) \equiv \lim_{T\rightarrow 0+}\limits S(Z,T)
   \;\quad
   \forall \; Z\; \in \Delta
\end{equation}

Assumption2 is assumption F of [12].It implies and is implied by the vanishing of the heat capacities
 $\,C_Z(T)\,$ (when the latter exist):
\begin{equation}
\lim_{T\rightarrow 0+}\limits \, C_Z (T) = 
   \lim_{T\rightarrow 0+}\limits \left[T\left(\frac{\partial S}{\partial
   T}\right)_Z
      \right] =  \lim_{T\rightarrow 0+}\limits \left[S (Z,T) - S (Z,0)\right] =0
\end{equation}
by (7). The continuity assumption (7) is standard: indeed that is how
the residual entropy $\,S_r\,$ obtained in [3] is compared with
experiment, yielding perhaps the best comparison between a theoretical
an experimental value in thermodynamics. See also section 2.

We shall refer to the set $\,\Gamma_0$
\setcounter{equation}{8}
\begin{equation}
\Gamma_0 = \{ T=0\} \times \Delta
\end{equation}
as the {\it zero temperature state space\/.} We may now state the

\vspace{0.3cm}

\noindent {\sc Third Law\/} \ \  The  zero temperature state
space $\,\Gamma_0\,$ is (physically) adiabatically inaccessible
from any point of the state space of the simple system.

\noindent The main consequence of the third law is:

\vspace{0.3cm}

\noindent{\sc Theorem 1.1 : Planck's Formulation of Nernst's Theorem\/}  \ \
 Under assumptions 1 and 2, the third
law implies that all points of $\,\Gamma_0\,$ are adiabatically
equivalent (see (2)). As a consequence, for a simple system in which the state space $\Gamma$ is parametrized by $(T,V,N)$ , where $V$ is the volume and $N$ the particle number, $S_r$ is a universal constant, independent of the specific volume $\frac{V}{N}$.

\vspace{0.3cm} 

\noindent{\sc Proof\/} \ \ By the statement of the third law and (3),
\[
S (Z,0) \ngtr S(Z', T_1) \quad \forall Z,Z'\; \in \; \Delta ,
      \forall T_1 >0\quad ,
\] 
and thus
$$
S(Z,0) \leq S(Z', T_1) \quad \forall T_1 >0\eqno{\rm (11a)}
$$

\noindent where $\, Z,Z'\,$ are arbitrary in $\, \Delta$. Taking
now, $\,T_1 \rightarrow 0+\,$ in (11a), and using (7), we find 
$$
S(Z,0) \leq S(Z',0) \eqno{\rm{(12a)}}
$$
Since $\,Z\,$ and $\,Z'\,$ are arbitrary in $\, \Delta $, we may
invert the roles of Z and $\,Z'\,$ in (11a) to obtain
$$
S(Z',0) \leq S(Z, T_1) \;\;\forall T_1 >0\eqno{\rm{(11b)}}
$$
and, again taking $\, T_1\rightarrow 0+\,$ in (11b) and using (7), we
get
$$
S (Z',0) \leq S(Z,0) \eqno{\rm {(12b)}}
$$
From (12a) and (12b)
\setcounter{equation}{12}
\begin{equation}
S(Z',0) = S(Z,0)
\end{equation}
which is the assertion of the theorem.
It is to be noted that (13) ammounts to Planck's restatement of the third law, which has been shown to be mandatory for homogeneous systems in reference [12].
By extensivity (4), (13) is scale
invariant. Thus we may restrict ourselves in (13) to scale invariant
variables $\,Z$, or, in other words, {\em intensive variables\/}
$\,Z$. In terms of the usual thermodynamic variables $(V,N,T)$, with $N$ the particle number and $V$ the volume, taking the scaling factor $t=\frac{1}{N}$ in (4), the residual entropy is written as $S_r=\frac{1}{N}S(\frac{V}{N},0)$.
By (13), $S_r$ is independent of the specific volume. This completes the proof.

\vspace{0.3cm}

\noindent{\sc Remark 1.1\/} \ \  Relation
\begin{equation}
S(Z,0) = S(Z', T_1) ,\;\; Z,Z'\; \in\; \Delta, T_1 >0
\end{equation}
is a possibility included in (11a). It means that the change $\, (Z', 
T_1) \rightarrow (Z,0)\,$ can only be performed in an idealized sense,
i.e., it takes infinite time, or an ``infinite number of
operations''. It leads directly to (13), upon taking the limit $\,T_1
\rightarrow 0+\,$ using (7).

\noindent We may thus state the third law in the following alternative
way: the zero temperature state space is unattainable in a finite time (or in
a finite number of steps). $\square$

We emphasize that (13) is not new (see, e,g., [13], p.2, for a recent
discussion). However, for instance in Landsberg's analysis ([14], [15]),
the unattainability law is seen to imply (13) only under a
thermodynamic stability assumption:
\begin{equation}
(\partial S/ \partial T)_Z >0 \qquad
   \forall \; T >0,\qquad\forall Z\,\in\,\Delta
\end{equation}
See also the discussion in [13], and [16] for a discussion of the third
law in a vein similar to [14], [15]: in [16] the authors {\em assume\/}
(14) (not stating that it is (part of) the thermodynamic
unattainability principle) and pretend to use (15) in an argument
which turns out to be circular. In our approach, (15) does not, as we
have seen, play any {\em explicit\/} role, although stability
conditions (concavity properties of the entropy) are essential for the
construction of [7]. 

On the other hand, in (7) there is implicit one of the asumptions in
([15], [16]), namely that
\begin{equation}
|S (Z,0)| <\infty
\end{equation}
We believe that(16) is, however, \textbf{universally} violated  in {\em
classical\/} statistical mechanics, a subject to which we now turn.

\vspace{1cm}

\noindent{\bf 2 -- Applications to Statistical Mechanics}
\vspace{0.3cm}

Let $H^Q_N (J)$ denote the Hamiltonian of a system of $N$ spins of spin quantum number $J$. The canonical quantum von Neumann entropy  $\,(\beta =1/kT)$ is 
$$ 
S^Q_N (B,T;J) = -k
\;Tr\;\left(\delta^Q_N \;\log \; \delta^Q_N\right) 
\eqno{\rm{(17a)}}
$$
with
$$
\delta^Q_N (J) = \frac{e^{-\beta H_N^Q (J)}}{Tr e^{-\beta H_N^Q (J)}}
\eqno{\rm{(17b)}} 
$$
and the classical entropy is 
$$
S^{Cl}_N  = - k \int d\Omega^N \;
\delta^{Cl}_N (\Omega^N)\log \; \delta^{Cl}_N (\Omega^N)
\eqno{\rm{(18a)}} 
$$
where
$$
\delta^{Cl}_N (\Omega^N) = \frac{e^{-\beta H_N^{Cl}}(\Omega^N)} 
  {\displaystyle\int d\Omega^N \,
  e^{-\beta H_N^{Cl} (\Omega^N)}}\eqno{\rm{(18b)}} 
$$
and $\, d\Omega^N =
  {\raisebox{-0.6em}{$\stackrel{\textstyle \otimes}
  {\scriptstyle i=1}$}}^{^{^{\!\!\!\!\!\!\!N}}} d\Omega^i =( sin \;\theta_i\;
  d\theta_i\;d\varphi_i)/4\pi$ is a \textbf{normalized} measure and
 $\,\Omega^N \equiv (\Omega_i)^N_{i=1},
  \;\Omega_i \equiv (\theta_i,\varphi_i)$.

By (17b)
$$
\delta^Q_N (J) \leq 1 \;\;\;\forall\;J < \infty\eqno{\rm{(19)}}
$$
which implies, by (17a)and the elementary inequality
$$
-x\log(x) \geq 0 \mbox{ if } x\in[0,1]
\eqno{\rm{(20)}}
$$
that 
$$
S_N^Q (B,T;J) \geq 0\;\;\;\forall J <\infty \eqno{\rm{(21)}}
$$
\vspace{0.3cm}

\noindent{\sc Proposition 2.1\/} ([9],Prop.1)
$$
S^{Cl}_N\le 0
\eqno{\rm{(22)}}
$$
\noindent{\sc Proof.\/}\\Follows from convexity of the function
$x\log(x)$($ x\ge0$):
$$
-x\log(x)\le 1-x
\eqno{\rm{(23)}}
$$

An immediate corollary of (21) and (22) is

\vspace{0.3cm}

\noindent{\sc Corollary 2.1\/}
$$
\lim_{J\rightarrow \infty}\limits \; \lim_{N\rightarrow \infty}\limits
   \frac{S^Q_N}{N} = \lim_{N\rightarrow \infty} \frac{1}{N}S_N^{Cl}
\eqno{\rm{(24)}}
$$

if the above limits exist.

By Lieb's theorem on the classical limit of quantum spin systems [10], the limit at the l.h.s. of (24) exists if $S^Q_N$ is replaced by the quantum free energy, and it equals the the limit on the r.h.s. of (24) with $S^{Cl}_N$ replaced by the classical free energy. The reason for (22) is that the inequality (20), viz., 

$$
{\delta^{Cl}_N} (\Omega^N) \leq 1 \eqno{\rm {(25)}}
$$

does not hold in general, because $d\Omega^N$ in (18b) is not a discrete measure. Indeed, (25) \textbf{must} be false at least for some values of $T$, otherwise (20) would yield a contradiction with (22) of proposition 2.1 by (18a).
This phenomenon is, of course,
well-known and has been remarked in this context ([17], 2.1.6.2  p.44,
2.26. pg. 57). In this connection it may be remarked that we may replace in (18b) $H^{Cl}_N(\Omega^N)$ by 
$$
H^{Cl}_N(\Omega^N)-E_0
\eqno{\rm{(26)}}
$$

where $E_0=\inf_{\Omega^N}H^{Cl}_N(\Omega^N)$ denotes the ground state energy of
$H^{Cl}_N$. Assuming the $d\Omega^N$ measure of the ground state configurations (26) to be zero leads us to guess, with the above-mentioned replacement, that 
$S^{Cl}_N$, given by (18a), tends to $-\infty$ as $T\to 0+$. Since, however, 
$d\Omega^N$ is a continuous measure, a rigorous proof of this assertion requires a detailed analysis of the $d\Omega^N$ measure of configurations
$\Omega^N$ such that $ H^{Cl}_N(\Omega^N)-E_0\le 1/\beta$ which, as far as we know, has not been done in general. We therefore content ourselves with the (very brief) analysis of two simple models which, moreover, throw light on the existence/nonexistence of the limits (24).
The first model  consists of $\,N\,$ noninteracting quantum
spins (of spin quantum number $\,J$) in an external magnetic field
along the $\,z\,$ -- axis, described by the Hamiltonian $\,(B>0)$:
$$
H^Q_N(J)=- B \sum^N_{i=1}\limits \left(\frac{S^z}{J} 
  +1\right) \eqno{\rm{(27a)}}
$$
where $\, S_i^z\,$ is the $\, z$-component of a spin $\,J\,$
quantum operator, together with the corresponding assembly of classical
rotors, with Hamilton function
$$
H^{Cl}_N (\vec{\theta}\,) = - B \sum^N_{i=1}\limits \left(\cos
\;\theta_i+ 1\right) \qquad \vec{\theta} \equiv
(\theta_i)^N_{i=1}\eqno{\rm{(27b)}}
$$
 where $\,\theta_i \in [0,\pi],
i=1,\ldots, N$, and the additive constant in (27) is arbitrary. The corresponding quantum entropy equals
\[
S_N^{Q} (B,T;J) = k\;\log 
  \left(e^{2N\beta B} e^{N\beta B/J} -1\right) \,-
\]
\[
-\, k\;\log \left(e^{NB\beta/J}-1\right) \,-
\]
\vspace{0.2cm}
\[
-k\beta \frac{(2NB+NB/J) e^{2N\beta B}
   e^{N\beta B/J}} 
     { e^{2N\beta B} e^{N\beta B/J} -1}
\, +
\]
\vspace{0.2cm}
$$
+\, k\beta NB/J\;\frac{e^{N\beta B/J}}
    {e^{2N\beta B/J} -1} \eqno{\rm{(27c)}}
$$

and the classical entropy is given by
$$
S^{Cl}(B,T)\equiv\frac{1}{N}S^{Cl}_N(B,T)=\;
-\frac{1}{2}k\beta B(\coth(\frac{\beta B}{2})+2)\;
  +\; k\;\log \;\frac{4\pi}{\beta B} + 
     \frac{3k\beta B}{2} + k\;\log
      \left(1-e^{-\beta B}\right)\eqno{\rm{(27d)}}
$$

from which:

$$
\lim_{J\rightarrow \infty}\limits \; \lim_{N\rightarrow \infty}\limits
   \frac{S^Q_N (B,T_i J)}{N} =0 \eqno{\rm {(27e)}}
$$
$$
\lim_{N\rightarrow \infty}\limits \; \lim_{J\rightarrow \infty}\limits
   \frac{S^Q_N (B,T_i J)}{N} = + \infty  \eqno{\rm {(27f)}}
$$

Clearly, (27e) and (27g) confirm the result of Corollary 2.1, but (27f) shows further that the limits $\, N\rightarrow \infty\,$ and
$\, J\rightarrow \infty\,$ of the entropy per spin in general {\em do
  not commute.\/}
In order to show that the phenomenon
exemplified in corollary 2.1 is not restricted to free
(noninteracting) systems, consider the quantum ferromagnetic
Heisenberg chain, with $\,\lambda >0$
$$
\widetilde{H}^Q_N (J) = - \frac{\lambda}{J^2} \, \sum^N_{i=1}\limits   
   \vec{S}_i \cdot \vec{S}_{i+1} \eqno{\rm {(28a)}}
$$
where $\,\vec{S}_i \equiv \left(S^x_i, S^y_i, S^z_i\right), i=1,
\ldots , N\;$ are (spin - $J$) quantum operators, $\, \vec{S}_{N+1} =
\vec{S}_1\,$ (periodic b.c.). The corresponding classical Heisenberg
Hamiltonian is 
$$
{\widetilde{H}}^{Cl}_N  = - \lambda \sum^N_{i=1}\limits 
    \vec{S}_i^{Cl} \cdot\vec{S}_{i+1}^{Cl} \eqno{\rm{(28b)}}
$$
with $\,\vec{S}_i^{Cl} \equiv (\sin \theta_i \;\cos \;\varphi_i, \sin
\,\theta_i\; \sin\;\varphi_i, \cos\;\theta_i), i=1, \ldots, N$. By
Joyce's result [18], 
$$
f(\beta) \equiv -\lim_{N\rightarrow \infty}\limits \,
   \frac{\beta^{-1}}{N} \,\log \,\widetilde{Z}^{Cl}_N = -\beta^{-1} \log
    [\sinh (\beta\lambda) / (\beta\lambda)]+\frac{1}{\beta}\log(4\pi) 
\eqno{\rm{(28c)}}
$$
where
$$
\widetilde{Z}^{Cl}_N = \int d \Omega^N \; e^{-\beta
   \widetilde{H}^{Cl}_N} \eqno{\rm{(28d)}}
$$

We have

$$
S^{Cl} (T) = \lim_{N\rightarrow \infty} \frac{1}{N} 
   S^{Cl}_N (T) = - \frac{\partial f}{\partial T} = 
   k\beta^2 \; \frac{\partial f}{\partial \beta} \eqno{\rm{(28e)}}
$$
by Griffiths' lemma [19], because $\,f\,$ is a concave function of
$\,T$, and, by (28e) and (28c),
$$
S^{Cl} (T) = k(1-\log{8\pi})-k\log(\beta\lambda) + c(\beta)
\eqno{\rm{(28f)}}
$$

with $c(\beta)\rightarrow 0$ as $\, \beta \rightarrow \infty$. We thus have:
\vspace{0.3cm}

\noindent{\sc Proposition 2.2\/} For the examples (27a,b) and (28a,b),
$$
\lim_{T\to0+}\lim_{N\to\infty}\frac{1}{N}S^Cl_N(T)=\;
\lim_{N\to\infty}\lim_{T\to0+}\frac{1}{N}S^Cl_N(T)=-\infty
\eqno{\rm{(29)}}
$$

The second line of (29) follows from (27d) and the analogue of (28f) for finite $N$. 

We make some remarks on particle systems. For
particle systems, the (physical) classical limit corresponds to the
statement that the difference between quantum and classical
(infinite-volume) free energies is small if the thermal wavelength $\,
\lambda = (2\pi \hbar^2 \beta/m)^{1/2}\,$ is small with respect to the
mean particle distance and to a characteristic length of the
potential, i.e., the physical parameter which is varied is the inverse
temperature $\,\beta \rightarrow 0$. See [20] for rigorous results
along this line. For the entropy, however, results are scarce: the
high energy limit of the microcanonical entropy is classical for a
class of systems, as shown in appendix of [20]. We thus expect that
for particle systems the classical limit of the entropy coincides with
the high temperature limit, and, therefore, propositions 2.1 and 2.2
do not have an immediate analogue. {\em If one extrapolates\/}, however,
the entropy of the classical ideal gas to $\,T\rightarrow 0+$, it is
well known that the analogue of (21) obtains (see [5] for
comments). It is also rigorously known from the seminal work of Lieb
and Yngvason that, in the case of Bosons, no semiclassical
approximation can be valid for the ground state even in the limit of
high dilution [21]. Finally, proposition 2.1 is also valid for particle systems
[9].

Two conclusions may be drawn from this section. Firstly, proposition 2.2 suggests that assumption 2 fails in an universal way for classical spin systems.
Secondly, corollary 2.1 shows that an {\em
  obstruction\/} exists in the classical limit regarding the entropy:
  the quantum result does not join smoothly to the classical one. This
  seems to be the real reason why the third law does not hold for
  classical systems, in the framework of statistical mechanics.

\vspace{1cm}
\noindent{\bf 3 --- Conclusion, Open Problems and Conjectures}
\vspace{0.3cm}

In this paper, we have formulated a precise version of the third
law. Our basic assumption is the continuity assumption~2, which is
expected to be valid for all systems obeying the laws of quantum
statistics. We conjecture that  classical systems do not obey (16) of assumption~2. Although a general proof is missing, we were able to illustrate the conjecture with two models, one of noninteracting, the other of interacting,
 classical spins in section 2.

The vanishing, as $T\to 0$, of derivatives of the entropy (when they exist) with respect to variables $Z\in\Delta$, e.g., 
$(\frac{\partial S}{\partial v})_T = (\frac{\partial v}{\partial T})_p$
(experimentally well confirmed, see [22], pp 58-62), does not follow from Theorem 1.1. A thermodynamic proof of such relations remains, thus, open. Theorem 1.1 suffices, however, to lend support to the physical picture implied by the third law (see also remark 1.1) in specific situations, e.g., the existence of a succession of ever decreasing steps in the adiabatic demagnetization of paramagnetic crystals ([22], Fig. 9.1).

We finally mention the controversial question of the third law for black holes, recently reviewed in [13]. In [6] some quantum models of so-called extremal black holes (see the following (35)) violating Theorem 1.1 were constructed, on the basis of which it was attempted to ``lay to rest the 'Nernst theorem' as a law of thermodynamics''. By ``thermodynamics'', in the previous quotation, it was meant ``conventional'' and not, specifically, black-hole thermodynamics. We should like to argue to the contrary, namely,that, in spite of having provided a rigorous foundation for Nernst's theorem, the problem of formulating the third law for (Kerr) black holes remains open.

In order to be more precise, we consider the general case of a (classical) Kerr-Newman black-hole of charge $\,Q$, angular momentum $\, J\, $ and mass $\, M$, which is supposed to describe the gravitational collapse of a rotating star,
and for which the thermodynamic identity becomes (see [23] and
references given there):
\setcounter{equation}{29}
\begin{equation}
TdS = dM - \Omega \, dJ - \Phi dQ
\end{equation}
where $\,\Phi\,$ denotes the electric potential.

We assume
$$
M² \geq a² + Q²\eqno{\rm (31a)}
$$
Let 
$$
a=J/M \eqno{\rm (31b)}
$$
and 
$$
r_\pm = M \pm \sqrt{M² - a² - Q²}
 \eqno{\rm (31c)}
$$
We have
$$
r_+ - r_- =2 \sqrt{M^2 - a^2-Q²}
\eqno{\rm(31d)}
$$
and the surface gravity
$$
\kappa = \frac{r_+ - r_ -}{2\alpha}\eqno{\rm (31e)}
$$
with
$$
\alpha = r^2_+ + a^2 = 2M² + 2M
   \sqrt{M² -a² -Q²} - Q^2\eqno{\rm(32a)}
$$
The area of the event horizon $\, r=r_+\,$  is given by
$$
A= 4\pi \alpha \eqno{\rm(32b)}
$$
We take for the entropy the Bekenstein-Hawking result $\, S_B\,$ (see, e.g.,[1]), which equals
$$
S_B = \frac{A}{4} = \pi \alpha \eqno{\rm (32c)}
$$

The angular velocity $\,\Omega\,$ of the black hole is
\setcounter{equation}{32}
\begin{equation}
\Omega = \frac{a}{\alpha} = \frac{a}{r_+^2 + a²}
\end{equation}

Our black hole radiates like a black body
of temperature
\begin{equation}
T = \frac{\kappa}{2\pi} = \frac{r_+ -r_-}{4\alpha\pi} =
   \frac{\sqrt{M^2 - a^2 -Q^2}}{2\pi \alpha}
\end{equation}
(the Bisognano--Wichmann--Hawking--Unruh effect, see [24] and references
given there: this a consequence of the behavior of quantum fields in the
presence of an event horizon).

The limit $\, T\rightarrow 0+ \,$ corresponds by (31) to {\em
  extremal\/} black holes, i.e., the equality
\begin{equation}
M²= a² + Q²
\end{equation}
holds in (31a). Physically, it represents the boundary of the instability region $M^2<a^2+Q^2$, which corresponds to very rapidly rotating bodies.
 For $T$ in a fixed, sufficiently small $(J,Q)$ neighborhood of $(0,0)$, it may be shown that one may solve the equation
\[
\left(\frac{\partial M}{\partial S_B}\right)_{J,Q} = 
   T (\neq 0)
\]
locally for $\, S_B$, to obtain $\, S_B = S_B(T,J,Q)$. It is on this
quantity that the limit $\,T\rightarrow 0+\,$ is to be performed,
which, by (34), (32a) and (32c), yields:
\[
\lim_{T\rightarrow 0+}\limits S_B(T,J,Q)= \pi 
   (2M (T=0)^2 - Q²)\,=
\]
$$
=\pi(2a^2+Q^2) = \pi \left(\frac{4J^2}{Q^2 + \sqrt{Q^4 + 4J^2}} + Q^2
   \right) > 0
\eqno{\rm{(36)}}
$$
whenever $\, (J,Q) \neq (0,0)$.

 This contradicts (13), because of the
dependence of the r.h.s. of (36) on the remaining variables
$\,(J,Q)$. Thus, the Kerr-Newman black hole provides an explicit
example in which assumption 2 holds but Theorem 1.1 is violated. This violation, commented in [6] for the special case $Q=0$, does NOT, however, contradict the unattainability of the zero temperature state (in the case of black holes, of the extremal region (35)) in finite time, because our derivation of Theorem 1.1 from the third law (as stated in the present paper) ALSO fails for (Kerr-Newman) black holes, for the following reasons:
a.) the rigorous thermodynamic framework of Lieb and Yngvason, on which we strongly relied, does not hold for black holes because extensivity (4) is not valid for black holes (and gravitating systems in general [17]);
b.) for black holes the second law in the form (3) does not hold for $S_B$, but rather for $S_B + S_M$ , where $S_M$ denotes the entropy of the total matter in the Universe (see, e.g., [1] and references given there). This is important, because $S_B$ may decrease in an adiabatic process.

It thus remains as a (challenging) open problem to devise for black holes a framework, analogous to Lieb and Yngvason's, from which the Beckenstein-Hawking entropy $S_B$ emerges as the entropy function, fulfilling the second and the still-to-be formulated third law.

We conclude with a few related conjectures. Due to the important role played by quantum fields in black hole physics, apparent from the notion of temperature (34), the above-mentioned framework should contemplate the thermodynamics of quantum fields, a subject still in its infancy [25] ( for a rigorous approach to the supposedly related D-brane states, see [26]). In particular, the vacuum fluctuations of the quantum fields at the horizon, leading to the short-distance divergence of the area-density of localization entropy  (a supposedly quantum version of $S_B$) [27], may lead to a ``discontinuity'' between $S_B$ and the still-to-be constructed quantum field theory from which black holes emerge in the classical limit, perhaps analogous to the ``obstruction'' found in section 2.

\vspace{1cm}

\noindent{\bf Acknowledgement:\/} We should like to thank Dr. Pedro
L.Ribeiro for helpful conversations , Dr. Francesco Belgiorno for several comments, and the referees, particularly referee 2, for crucial comments and corrections. This work was partially supported by CNPq and FAPESP (Brazil).

\pagebreak

\noindent{\bf REFERENCES}

\begin{list}{}{\setlength{\leftmargin}{7mm}\labelwidth2.5cm
\itemsep0pt \parsep0pt}
\item[{[1]}] {\sc R.M. Wald\/} -- Black Holes and Thermodynamics -- ar
  Xiv: gr -qc / 970202201 (11-2-1997) -- proceedings on the Symposium
  on Black Holes and Relativistic Stars, Chicago, 1996.

\item[{[2]}] {\sc M. Aizenman\/} and {\sc E.H. Lieb\/} -- {\sl J.
    Stat. Phys.\/} {\bf 24}, 279 (1981).

\item[{[3]}] {\sc E.H.Lieb\/} -- {\sl Phys. Rev. Lett.\/} {\bf 18}, 692
  (1967).

\item[{[4]}] {\sc E. Fermi\/} -- Notes on Thermodynamics and
  Statistics. University of Chicago Press, 1966.

\item[{[5]}] {\sc W. Pauli\/} -- Lectures on Physics -- vol 4 -- {\em
    Statistical Mechanics} -- ed. by C.P. Enz, The MIT Press, 1973.

\item[{[6]}] {\sc R.M. Wald\/} -- {\sl Phys. Rev.\/}, {\bf D56}, 6467
  (1997).

\item[{[7]}] {\sc E.H. Lieb\/} and {\sc J. Yngvason\/} -- {\sl Phys.
      Rep.\/} {\bf 310}, 1--96 (1999) -- Erratum {\sl Phys. Rep.\/}
    {\bf 314}, 669 (1999).

\item[{[8]}] {\sc C. Nisoli et al\/} -- {\sl Phys.Rev.Lett.\/} {\bf 98},
217203 (2007).

\item[{[9]}] {\sc D.W.Robinson\/} and {\sc D. Ruelle\/} -- {\sl Comm.
     Math. Phys.\/} {\bf 5}, 5 (1967).

\item[{[10]}] {\sc E.H. Lieb\/} -- {\sl Comm. Math. Phys.\/} {\bf
      31}, 327 (1973).

\item[{[11]}] {\sc W. Abou Salem\/} and {\sc J. Fröhlich\/} -- Status
  of the Fundamental Laws of Thermodynamics -- arXiv math-ph/0604067;
  {\sl Lett. Math. Phys.\/} {\bf 72}, 153 (2005).

\item[{[12]}] {\sc F. Belgiorno\/} -- {\sl J.Phys.A\/} {\bf 36}, 
  8165 (2003).

\item[{[13]}] {\sc F. Belgiorno\/} and {\sc M. Martellini} -- {\sl Int.
      Jour.Mod.Phys.\/} {\bf D13}, 739 (2004).

\item[{[14]}] {\sc P.T. Landsberg\/} -- Thermodynamics with quantum
  statistical illustrations (Interscience Publishers, N.Y. 1961).

\item[{[15]}] {\sc P.T. Landsberg\/} -- Thermodynamics and Statistical
  Mechanics -- Dover, N.Y. 1990.

\item[{[16]}] {\sc D. ter Haar\/} and {\sc H. Wergeland\/} --
    Elements of Thermodynamics -- Addison Wesley 1966.

\item[{[17]}] {\sc W. Thirring\/} -- Lehrbuch der Mathematischen Physik
  Bd. 4 -- Quantenmechanik grosser Systeme - Springer 1980.

\item[{[18]}] {\sc G.S.Joyce\/} -- {\sl Phys. Rev.\/} {\bf 155}, 478
  (1967).

\item[{[19]}] {\sc R.B. Griffiths\/} -- {\sl J. Math. Phys.\/} {\bf
  5}, 1215 (1964).

\item [{[20]}] {\sc W. Wreszinski\/} and {\sc G. Scharf\/} -- {\sl
  Comm. Math. Phys.\/} {\bf 110}, 1 (1987).

\item[{[21]}]{\sc E. Lieb \/} and {\sc J. Yngvason\/} -- {\sl
  Phys. Rev. Lett.\/} {\bf 80}, 2504 (1998).

\item[{[22]}] {\sc F. Pobbell \/} -- Matter and Methods at Low 
  Temperatures - Third edition - Springer 2007 
 
\item[{[23]}] {\sc E. Abdalla, W. Qiu, B. Wang\/} and {\sc R.K. Su\/}
  -- {\sl Phys. Rev. } {\bf D64}, 27506 (2001).

\item[{[24]}] {\sc G.L. Sewell\/} -- {\sl Ann. Phys.\/} {\bf 141},
      201 (1982).

\item[{[25]}] {\sc C. Jäkel\/} -- Thermal Quantum Field Theory --
  Encyclopaedia of Mathematics {\bf 89}, Elsevier 2006.

\item[{[26]}] {\sc P.L. Ribeiro\/} --  arXiv:0712.0401(math-ph).
  
\item[{[27]}] {\sc B. Schroer\/} -- Class. Quan. Grav. {\bf 23}, 5227
  (2006).

\end{list}

\end{document}